# Chaos spectrum - semiconductor laser with delayed optical feedback

**D M K**ANE[1,2*] **AND M R**ADZIUNAS[3,2]

[1] *Research School of Physics, The Australian National University, Canberra, ACT 2600, Australia*
[2] *MQ Photonics Research Centre, and, Dept. of Physics and Astronomy, Macquarie University, Sydney, NSW 2109, Australia*
[3] *Weierstrass Institute for Applied Analysis and Stochastics, Mohrenstrasse 39, 10117 Berlin, Germany*
*[deb.kane@anu.edu.au](deb.kane@anu.edu.au)*

**Abstract:** Maximizing the rf bandwidth associated with the chaotic output from tailored operation of nonlinear semiconductor laser systems is an ongoing research effort. The early pioneering research was done in semiconductor laser with delayed optical feedback systems, which continue to be researched. We report numerical simulations of this system, using a travelling wave model. The results provide new insights into the impact of key device parameters affecting the chaos bandwidth and spectrum. Linewidth enhancement factor and the nonlinear gain saturation parameter are found to be the most important parameters when seeking to optimize the chaotic output. We reassess a standard definition being used to report chaos bandwidth. We propose that more spectral information should be reported if numerical and experimental research results are to be of most value into the future. A database from previous experimental study is also analyzed to connect with the predictions of the numerical simulations. This elucidates the links between the chaos bandwidth achieved in real systems and the semiconductor laser parameters. The results inform recommendations for semiconductor laser parameters that will better support broadband chaos generation in whatever semiconductor-gain-medium-based nonlinear system approach is being used. They elucidate the physics of both the envelope and the fine structure of the rf spectrum of coherence collapse.

## 1. Introduction

The system of a semiconductor laser (SL) with delayed optical feedback (DOF, see fig. 1) has been researched extensively since the mid 1980's, and has been the subject, or a major part, of several reviews and monographs [1-4]. This attention has arisen because the system is an excellent testbed and exemplar of nonlinear dynamics in general, and nonlinear laser dynamics in particular. Also, the SLDOF has been researched for a range of technological applications [5], such as secure communication based on synchronization between a chaotic transmitter and a slave receiver (e.g. [1-5]), random number generation (e.g. [3-5]), key distribution and exchange (e.g. [6]), and information processing including photonic reservoir computing (e.g. [7-12]). For these applications, chaotic output, one of the dynamical regimes that the system can be operated in, is generally required. The radio frequency (rf) bandwidth of the chaos, typically several GHz to several tens of GHz, has been a key characteristic of the system output that researchers have sought to define appropriately, maximize and optimize. The complexity of the chaos is also important. Suppression of any signature of the delay time ($\tau = c/2L$, where $L$ is the length of the external cavity (EC) and $c$ is the speed of light in free space) is required for applications such as secure communication. When seeking to achieve ever increasing chaos bandwidth which is as flat as possible over the rf bandwidth, there are additional technological approaches that can be employed. Examples include photonic integrated devices (e.g. ([11-14]), systems involving both optical injection and delayed optical feedback (e.g. [10, 15, 16]), or multiple devices with mutual coupling (e.g. [17, 18]). Use of additional processes to actively suppress the time delay signature is also sometimes applied [19, 20]. There is a plethora of such systems reported in the literature. They use different types of SLs (Fabry Perot (FP) lasers, distributed feedback lasers, vertical cavity surface emitting lasers (VCSELs), quantum dot lasers, quantum cascade lasers, semiconductor optical amplifiers in ring configurations), and, one or more of the standard approaches for destabilizing the laser.

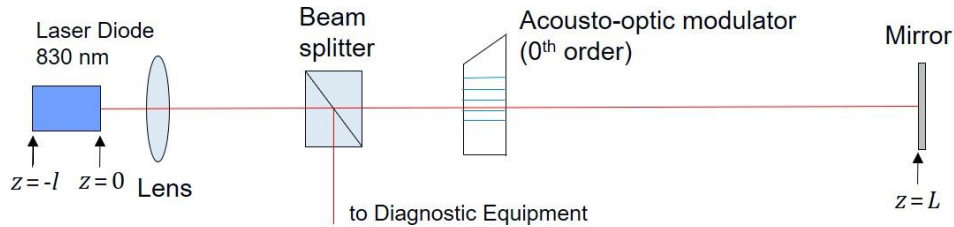

Fig. 1. Schematic of a SLDOF system. $z = L$ is for an empty EC case. The optical pathlength of the EC is the relevant length for a real system which will normally include optical elements like the collimating lens, acousto-optic modulator.

Despite the enormous amount of research on SLDOF systems, the literature lacks an account of the broadband chaos from a bulk-optic, free-space-propagation system using a FP SL, covering the full range of optical feedback. SLDOF systems were the progenitor from which the state-of-the-art devices and systems for broadband chaos generation have descended. It is probably the absence of high bandwidth photodetectors and real-time oscilloscopes, and, also the limited capacity to generate and analyze big data sets, at the time of the pioneering SLDOF studies that has led to this omission to date. Broader bandwidth chaos has been demonstrated from systems using different SL devices and/or different methods of inducing the chaotic output. However, a state-of-the-art numerical study of the chaos spectrum of the SLDOF using a standard FP SL gives insights that are relevant to such other broadband chaos systems as well. Also, we revisit the definition of chaos bandwidth. We encourage researchers to report more detailed characterization of the spectral output of chaotic lasers. More detailed information will be more useful to future researchers and developers when considering a given system for a given application.

In the subsections of the introduction which follow we remind the reader of the early description of optical feedback in a FP SL We introduce key features of the rf spectrum of the chaotic output from a SLDOF; the travelling wave model and the visualizations used to report results from the model. Also the experimental system and the publicly available experimental datasets that are used, with appropriate analysis, to make comparison with the simulations are introduced.

## 1.1 Conceptual background

It is more than 50 years since dynamical effects were reported in a SL coupled to an EC [21]. The close connection between the GHz bandwidth dynamics and the spectra, both optical and rf, was key to experimental study of this system in the first 30-35 years. The subsequent availability of real-time oscilloscopes with measurement bandwidths of order 10 GHz to a few tens of GHz, combined with computer control and large dataset collection, has enabled high density mapping of the dynamics as a function of relevant variables of the SLDOF system (for example [22]). It was observed early that enhancement of the relaxation oscillation of the SL was associated with destabilization by delayed optical feedback. But, also suppression of relaxation oscillations was observed when strong optical feedback achieved single frequency, narrow-linewidth output from this system [23]. The latter is the basis of the single frequency, tunable SL systems that are ubiquitous in applications today. A seminal step in theoretical understanding of this systematic range of different output dynamics with changing optical feedback level was made by Lang and Kobayashi – the L-K model [24]. It models the excess number (or density) of carriers with respect to the threshold value and the complex slowly varying amplitude of the electric field for a single longitudinal and transverse mode FP SL, in the rate equation limit. It includes the influence of the optical feedback on the electric field, in general complex-valued, through a time delayed field term with feedback rate constant η [1, 2, 24-26]. The L-K model, and its further developments, predict much of the complex behavior that is observed experimentally. The exception is that the model is applicable to low levels of optical feedback and ignores the presence of multiple longitudinal modes (resonances) in the solitary laser. It neglects higher-order feedback terms due to multiple roundtrips of the field in the EC and therefore does not cover some observations at high optical feedback levels such as in [22, 23]. Moreover, the L-K model is not suited for modeling of SLDOF systems with small or vanishing field reflection at the front facet of the FP diode facing the EC. In contrast, the traveling wave (TW) model used in this work does extend to arbitrarily high optical feedback levels and naturally accounts for spatial distributions of fields and carriers, field reflections at the SL facets, and multiple longitudinal modes of a solitary, free running FP SL.

The term coherence collapse was first used in 1985 [27] to describe the sudden transition from single frequency operation to a multiple EC mode output with a broadband rf spectrum. This transition was observed for a negligible change in optical feedback level. This was soon linked with systematic experiments that defined and mapped the regimes of dynamics for the SLDOF system, as a function of EC length and optical feedback level [28, fig. 8]. The regime I-V classification from this research is still in use today. The dramatic transition from regime III to IV is coherence collapse and regime IV is called the coherence collapse region of the dynamics which has also been shown to be chaotic. At high optical feedback levels, only readily achieved using a SL with one high reflectance facet and an AR or low reflectance coating on the device facet facing the EC, another sudden transition from coherence collapse to narrow linewidth, single frequency operation is observed. This is the regime IV to V transition. A more recent update of the regime I-V mapping has been published [29] which shows features of the dynamics from the SLDOF system that have been elucidated since the first version was published. Overall, the system is a very complex one, but it is the two transitions into and out of coherence collapse, the extent of the coherence collapse region (regime IV), and the systematic variation in the optical spectrum and the rf spectrum within the coherence collapse region that are the primary foci of this study. From the enormous literature describing the SLDOF in detail (for example [1-4, 22-29]) we have found there is a relatively small subset of the theory that is required to give the key insights for interpreting the numerical simulations presented here-in. These in turn inform making connections with experimental results. Recommendations of preferred laser parameter values, that will support broadband chaos generation, result from the simulations. The L-K model presentation of ([1] ch. 2, or [2] ch. 4) is used to inform the following.

Let us assume that $F_e(t)$ and $F_i(t)$ are the SL-emitted, and the reinjected complex slowly varying optical field amplitudes, respectively, on the EC-side, related by eqn. (1.1). Here κ is the fraction of the emitted optical field amplitude that is reinjected, and φ is the frequency independent field phase shift within the EC. τ is the field roundtrip time in the EC. For FP SLs, which are considered in this paper, the complex feedback factor $\kappa e^{i\varphi}$ is related to the feedback rate η used in the non-scaled L-K model by eqn. (1.2). $\tau_0$ denotes the field roundtrip time in the FP SL and $r_0$ is the complex amplitude reflectance at the front facet (z=0 in fig. 1). The consideration of the phase factor φ can be of importance when investigating an initial destabilization of an SL with an EC of up to a few millimeters length. Here we note, however, that the L-K modeling approach is better suited for consideration of the SLs with small amounts of feedback with long delay. Long delays are implemented in most experimental free space propagation SLDOF systems (eg. [22]). According to the L-K model, the optical frequencies $\omega_s$ of the steady state solutions in the presence of optical feedback, are given by eqn. (1.3). $\omega_0$ is the optical frequency of the solitary FP SL, assumed to be single mode,. $\alpha_H$ is the linewidth enhancement factor (LEF), which describes the enhanced phase change caused by a change in gain in SLs, as compared to other common lasers. One way this manifests is as a free-running linewidth for a SL chip which is enhanced by a factor of $(1 + \alpha_H^2)$ compared to the standard Schawlow Townes linewidth.

$$F_i(t) = \kappa e^{i\varphi} F_e(t - \tau), \qquad \text{eqn. (1.1)}$$

$$\eta = \frac{1-|r_0|^2}{r_0 \tau_0} \kappa e^{i\varphi}, \qquad \text{eqn. (1.2)}$$

$$\omega_s = \omega_0 - |\eta|\sqrt{1 + \alpha_H^2} \sin(\omega_s \tau - arg(\eta) + arctan(\alpha_H)). \qquad \text{eqn. (1.3)}$$

These equations, and their subsequent development, have been used in a rich literature on this complex system (for example [1-4, 22-29]). Only the salient and simplified features need be documented here. The number of steady state, fixed point solutions of eqn. (1.3), known as external cavity modes (ECMs), increases with $|\eta|$. Among ECMs one typically distinguishes possible stable *modes* and *anti-modes*. The latter are unstable saddle-type steady states with an odd dimension of the unstable manifold. The number of steady state ECM solutions that will occur for different system settings are described by 2D maps [26]. These start from one mode and show an overarching trend of increasing in number, by mode-antimode pairs being added as $|\eta|$ increases. After the critical feedback level for coherence collapse is reached, increasing it further, leads to a quickly increasing number of ECMs lasing and broadening. More importantly the optical frequency span within which the steady state solution ECMs occur, scales as $2|\eta|\sqrt{1 + \alpha_H^2}$, as is seen in ECM ellipse representations [1-4]. Also, this can be conceptualized as the instantaneous frequency from the SLDOF sweeping repeatedly through a range, ending at about the free running $\omega_0$, and with a lower bound that decreases with the level of the optical feedback, scaling proportional to $|\eta|\sqrt{1 + \alpha_H^2}$. On average, the instantaneous optical frequency is spending more time in sub-bands centered on the ECM frequencies. Such behavior has been indicated by a population inversion versus phase-difference-per-round-trip visualization from L-K modeling for a SLDOF system operating between low frequency fluctuation dynamics and stably on the maximal gain mode (MGM) [30, fig. 2 inset] having a frequency offset $\omega_0 - \omega_{MGM} \approx |\eta| \alpha_H$, which for large LEF is just a bit smaller than $|\eta|\sqrt{1 + \alpha_H^2}$. A DFB laser was used in [30].

The relaxation oscillation, and its associated damping rate, are important parameters for determining the specification of the output of the SLDOF system throughout the coherence collapse region (see, for example, [1-4, 24-29], in particular [25]). The route to chaos in the SLDOF system is initiated by undamping of the relaxation oscillation. Theory of the relaxation oscillation frequency (ROF) has reported it will change with optical feedback level [2], but calculated results using that theory have not been presented. The theory suggests the ROF can either decrease or increase with increasing optical feedback depending on the details [2]. Experimental measurements have been reported for low levels of optical feedback that includes three data points just inside the coherence collapse region close to the regime III-IV transition [26]. The experiment showed a small reduction in the ROF of at most 1% for an increase in feedback fraction from ~3.3% to ~4% [26]. The numerical simulations that will be presented here allow the role of relaxation oscillations in the transition into, throughout, and, the transition out of the coherence collapse region to be more clearly articulated. These show the dependence of the frequency of the ROs on optical feedback level, in these regions of the dynamics, is weak.

*1.2 The traveling wave model*

For simulations of the spatiotemporal dynamics in the FP type diode laser a previously introduced traveling wave (TW) model is used [31-33]. It is based on the partial differential equations describing the longitudinal (*z*-dimension of space, see fig. 1) and temporal evolution of the complex slowly varying counter-propagating optical fields, $E_+(z,t)$ and $E_-(z,t)$ in the FP laser ($z \in [-l, 0]$):

$$\left(\frac{n_g}{c}\partial_t \pm \partial_z\right) E_\pm(z,t) = -i(\beta(N, \varepsilon|E|^2) - iD)E_\pm(z,t) + F_{sp}^\pm. \qquad \text{eqn. (1.4)}$$

The reflection and transmission boundary conditions at $z = -l$ and $z = 0$ [31, 32] that must be fulfilled are:

$$E_+(-l,t) = -r_{-l}^* E_-(-l,t), \quad \begin{cases} F_e(t) = \sqrt{1-|r_0|^2} E_+(0,t) - r_0^* F_i(t) \\ E_-(0,t) = \sqrt{1-|r_0|^2} F_i(t) + r_0 E_+(0,t) \end{cases}. \qquad \text{eqn. (1.5)}$$

The emitted and reinjected fields $F_e(t)$ and $F_i(t)$ are related by eqn. (1.1). $\beta = i\frac{g_T-\alpha_0}{2} - \frac{\alpha_H g'(N-N_{tr})}{2}$, with $g_T = \frac{g'(N-N_{tr})}{1+\varepsilon|E|^2}$, is the propagation factor, which depends on the carrier density $N(z,t)$ governed by the carrier rate equation

$$\frac{d}{dt}N(z,t) = \frac{I}{qV} + \frac{U_F'}{qVR_s}(\bar{N} - N) - \frac{N}{\tau_N} - \frac{c}{n_g}\Re\sum_{v=\pm} E_v^* \cdot (g_T(N, |E|^2) - 2D)E_v, \qquad \text{eqn. (1.6)}$$

and on the local photon density distribution $|E|^2=|E_+|^2+|E_-|^2$. The model naturally accounts for multiple longitudinal modes (resonances) of the SL [31], which is in contrast to a standard L-K-modelling approach. For limiting the number of these modes, a linear operator $D$,

$$DE_\pm = \frac{g_p}{2}(E_\pm - p_\pm), \qquad \frac{\lambda_0^2}{2\pi c}\frac{d}{dt}p_\pm = \frac{\gamma_p}{2}(E_\pm - p_\pm) - i\lambda_p p_\pm, \qquad \text{eqn. (1.7)}$$

which fits the material gain dispersion to a Lorentzian function of finite width $\gamma_p$, is used [31]. Further descriptions of the TW model are also given in [34], for example. This model has been used previously to complete a study of how the SLDOF system output changes, as a function of the level of the optical feedback, when the coherence of the optical feedback field is systematically varied [32]. In the present study the optical feedback is treated as being fully coherent. The focus is on the mappings showing how the optical frequency spectrum and rf spectrum change with the optical feedback level, κ. The linewidth enhancement factor (LEF) $\alpha_H$ and the nonlinear gain compression factor $\varepsilon$ are the two parameters found to sensitively affect these spectra of the output.

$F_{sp}$, $c$, $q$ in the equations above are Langevin noise sources, speed of light in vacuum, and electron charge. $\bar{N}$ denotes the spatial average of $N$, whereas the remaining parameters are briefly introduced in Table 1. Many of the values for the parameters used in this study have been translated from [35] where they were given to model an APL-830-40 FP SL from Access Pacific Ltd.

## 1.3 Chaos bandwidth and spectrum

Over the years there have been a number of definitions put forward and applied for measuring the chaos bandwidth from the rf spectrum of the chaotic output of laser systems operating in the coherence collapse regime. The definition that has been accepted and applied by most researchers in recent times defines the chaos bandwidth as the frequency range, measured from dc, that accounts for 80% of the total rf power [36]. As a single measure this is clearly defined and can be consistently applied to facilitate useful comparisons between systems. But it is timely to reflect on the standard definition of bandwidth in electronic contexts. When giving the specification for a high-end amplifier, for example, the 3dB bandwidth is defined by the low and high frequencies at which the gain has dropped by 3dB compared to its maximum value, which is required to be flat to within 1 dB within the central span. The specification is dictated by the requirements of the application of the amplifier, for example for use in a high-fidelity sound system. Many of the applications for chaotic lasers are future applications that have not yet been scoped. As such, the specification for the chaos bandwidth and spectrum for a given future application is uncertain or unknown. This leads to the need to specify chaotic output more fully, in ongoing research, than just to report the dc to 80% total rf power bandwidth. Even defining the 80% total rf power differently, such as taking the central portion of the frequency span with 80% of the total power, or, summing up only the strongest discrete spectral segments which account for 80% of the total power, leads to estimates of the chaos bandwidth that are reduced by up to an order of magnitude [37].

Table 1. Constants and parameter values used in the simulations and/or estimated for the APL SL

| Parameter | Meaning | Value |
|---|---|---|
| $N_{tr}$ | transparency carrier density | $10^{24}$ m$^{-3}$ |
| $g'$ | differential gain | $1.036 \times 10^{-20}$ m$^2$ |
| $\lambda_0$ | center wavelength | 830 nm |
| $n_g$ | group velocity index | 3.7 |
| $\alpha_0$ | field losses | 6000 m$^{-1}$ |
| $\alpha_H$ | linewidth enhancement factor | 3.5 (and varied) |
| $\varepsilon$ | nonlinear gain compression factor | $3 \times 10^{-23}$ m$^3$ (and varied) |
| $V$ | volume of active region | 300 μm × 5 μm × 0.1 μm |
| $r_0$ | amplitude reflectance at $z=0$ facet | $(0.05)^{1/2}$ |
| $r_{-l}$ | amplitude reflectance at $z=-l$ facet | $(0.95)^{1/2}$ |
| $\tau_N$ | carrier lifetime | 2 ns |
| $\tau$ | delay time in external cavity | 4.5 ns |
| $\tau_0$ | propagation time in the FP SL | 7.4 ps |
| $\lambda_p$ | detuning of gain peak from $\lambda_0$ | 0 nm |
| $\gamma_p$ | FWHM of Lorentzian gain | 30 nm |
| $g_p$ | amplitude of Lorentzian gain | $10^4$ m$^{-1}$ |
| $I$ | injection current | 55 mA |
| $I_{th}$ | threshold current | ~25 mA |
| $R_s$ | series resistance | 1 Ω |
| $U'_F$ | derivative of the Fermi level separation | $3.5 \times 10^{-26}$ Vm$^3$ |
| $\kappa$ | feedback amplitude (part of emission) | varied between 0 and 0.3 |
| $\varphi$ | feedback phase | 0 |

The rf spectrum is also commonly shown on a dB scale which de-emphasizes the spectral components on the scale of the external cavity mode spacing which occur in the chaotic SLDOF. Figure 2 illustrates these issues schematically. The three rf spectra, which could correspond to a chaotic output, are shown (not to scale). They might all have a common 80%-rf-power-bandwidth but they have very different spectral content and would not all, necessarily, meet the needs of a given application for a chaotic output. We recommend that publishing the full rf spectrum data, or a time series of output power from which it can be calculated, is the most appropriate way for publications to future-proof their findings for later use. It is also recommended that a more complete description of the key spectral features be given in text.

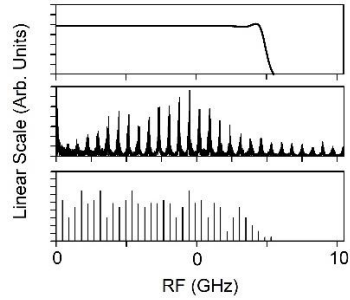

Fig. 2. Contrasting schematics of different rf spectra with a common 80% total rf power bandwidth

### *1.4 Experimental system datasets used for comparison with simulations*

The simulations with the TW model aim to connect with a publicly available dataset from an experimental SLDOF system. Fig. 1 is a schematic of the system which uses a multiple quantum well, 830 nm, FP SL (model APL-830-40 FP, Access Pacific Ltd). This system, the datasets it has generated, and the science that has been already elucidated using the data are described in a number of publications, for example [22, 38-40]. SL injection is controlled by directly varying the injection current to the device (Profile ITC502 Laser Diode Combi-controller). The laser was held at a constant temperature of 25°C. The system used a high reflectance external cavity mirror to provide the optical feedback. The laser output beam was collimated with an 8 mm focal length aspheric lens and then passed through a 50:50 cube beamsplitter. An acousto-optic modulator (AOM) (G&H 23080, Gooch and Housego) was used to control the fraction that was reflected back into the laser. The zeroth order beam of the AOM is used for the optical feedback. The beam splitter directs half the laser output into a 22 GHz detection bandwidth fast photodiode (DCS30S 22 GHz, Discovery Semiconductors).

Two datasets are used in this study. The first is referred to as the "4 GHz dataset" and it was collected sampling the photodiode signal at 20 GSamples/s for 1 µs (sampling interval 50 ps) using a digital storage oscilloscope (Agilent Infiniium 54854A DSO) with a 4 GHz real-time bandwidth. Output power time series were collected while stepping through values of the laser injection current and the fraction of delayed optical feedback. The injection current was swept from 45 mA to 70 mA in 0.1 mA steps. The zeroth order transmission of the AOM, which is proportional to the optical feedback level, was adjusted in 351 non-uniform steps between 75.5% and 6.5% (voltage to the AOM ranging from 0.80 V to 0.45 V in 0.01 V steps). Taking the double pass through the AOM and the beamsplitter into account, along with an estimate of 25 % for the coupling efficiency of the light back into the SL, the actual feedback percentage is reduced to estimates of order 10% to 0.5%. The corresponding range of amplitude feedback factors $\kappa$ is 0.32 to 0.07. Each of the 88,101 time series in the dataset consist of 20,000 points.

The second is the "16 GHz dataset". It contains 50,451 output power time series recorded using a 16 GHz real-time oscilloscope (Tektronix DSA72004C). The injection current has 251 values: 45 mA to 70 mA in 0.1 mA steps. The voltage to the AOM which controls the optical feedback has 201 values from 0.4 V to 0.8 V in 0.002 V steps. Higher voltage corresponds to lower optical feedback as more of the return beam is diverted into the first order beam of the AOM which is separated by angle and blocked. The corresponding range of intensity feedback, taking coupling efficiency and output coupling into account, is about 10% to 0.5%. Each time series contains 50,000 values of photodiode voltage measured across the 50 Ω oscilloscope input, sampled at 50 GSamples/s (20 ps per data point) for a total record length of 1 microsecond.

## 2. Results and discussion

## 2.1 Numerical simulations - dB scale

The numerical simulations were carried out using laser and system parameters listed in Table 1, unless stated to be different. The simulations presented are for a single injection current of 55 mA, about 2.2 times the threshold injection current. Fig. 3(a) shows power of the rf spectrum of the output of the SLDOF system using a color coded dB scale, as a function of the optical feedback factor κ, for the frequency range 0-160 GHz. The longitudinal mode spacing of the FP SL is 0.31 nm or ~135 GHz. The lower part of fig, 3(a) (0 - ~70 GHz) shows the rf spectrum associated with the intensity noise/chaos of the individual longitudinal modes. The upper part (~70 – 160 GHz) shows the rf components generated by nonlinear mixing of nearest neighbor longitudinal modes. It is the rf spectrum in the simulations up to the detection bandwidth of a specific experiment that can be compared with the experiment. This is a limited detection bandwidth of typically 15-50 GHz. Thus, the simulations give rf spectral information over a larger range than has been the subject of measurement to date. Over the full 0-160 GHz frequency range of the simulations the background level of the rf spectrum of the SL with small feedback is about 5 dB until coherence collapse occurs at a feedback value of ~0.011. Considering the frequency range 0 - ~50 GHz, this section of the rf spectrum then shows a peak about the ROF broadening both to higher and lower frequencies, with increasing feedback factor. The ROF is ~5 GHz for the simulation conditions of fig. 3. The peak associated with the ROF may flatten as feedback is increased. There is a second main peak in the rf spectrum. This second peak occurs at a higher frequency with increasing κ. It is at ~15GHz at κ ~0.20 in fig. 3(a). We refer to this spectral frequency as the primary swept-to-frequency (StF). The rf power level is also elevated between these two peaks in coherence collapse. At feedback of ~0.20 the system restabilizes. This represents a transition from coherence collapse to nearly single frequency operation, the regime IV to V transition. A third significantly weaker rf peak observed in Fig. 3(a)is seen at ~45 GHz at $κ$ ~0.20 and is always about three times the StF for a given $κ$. This is a secondary StF rf peak which corresponds to the frequency or wavelength separation of the solitary-FP-SL-resonance and the location of the steady state with the smallest possible threshold carrier density. The steady states of the TW model for the spatially-extended compound cavity consisting of the SL and EC (compound cavity modes, CCMs) were studied in, for example, [31,33,41]. They are TW model analogs of ECMs of the L-K model, whereas the smallest threshold state is known as a maximal gain mode in L-K and TW models. For feedback factors greater than 0.20 the system restabilizes and destabilizes, repeatedly, over small ranges of κ in the simulations. In this region, the calculated carrier density shows only a small-range-variation, whereas the rf power at any given frequency is at least 10 dB lower than it is for the main coherence collapse section of the dynamics. The optical spectrum is close to single frequency for these transitions. We note that at least some of these regimes with slightly enhanced spectral peaks represent dynamic transients toward more regular states. The 1μs time window used in the simulations at each new value of κ may have been insufficient for completing a switch between two single-CCM-defined states with only a tiny difference in their threshold carrier densities.

Fig. 3(b) shows the optical spectrum as a function of κ, with a color-coded dB power scale. The spectra are normalized such that the maximum of each calculated spectrum is assigned a 0dB value. Note that the blue colors represent 20 dB or more of mode suppression relative to the highest power (darkest red in the figure). For small κ, the system operates at the single wavelength (relative wavelength value of ~0.5 nm)-close to a resonance of the solitary SL, until coherence collapse occurs at a κ value of ~0.011. Seven solitary SL resonances then contribute to the dynamics in the wavelength range shown, after coherence collapse. Emission in the vicinity of four of these resonances has at least 10 dB more power than the weaker ones. The wavelength of the major spectral peak in the surrounding of each FP resonance is close to the location of the corresponding MGM wavelength. It can be useful to describe this as the swept-to-wavelength to emphasize its role as the upper boundary of wavelength for the dynamical output. As κ increases the spectra in Fig 3(b) show the wavelength sweeping to MG mode wavelengths but making rapid excursions back to, and even slightly shorter than the resonance wavelengths of the solitary FP laser. (More detailed representations of this scenario are shown in Fig. 4, Fig. 4(d) in particular). The coverage of this wavelength range is more complete for κ values up to κ ~ 0.12. For still higher values of κ most of the power occurs, on average, at wavelengths closer to the MGM (swept to) wavelengths.

A simplified TW model, which neglects gain dispersion, nonlinear compression, and the spatial distribution of carriers, has been used to estimate the position of the MGM in the vicinity of the chosen solitary SL resonance. Following [33,41], for the considered SLDOF we can derive the equation

$$F(\Delta\omega, \Delta G) = \cos(\tau_0 \Delta\omega - \alpha_H l \Delta G) - \frac{\cosh(l\Delta G) - \kappa^2 \cosh(l\Delta G - \ln|r_0|^2)}{1-\kappa^2} = 0, \qquad \text{eqn. (2.1)}$$

representing positions of all CCMs for the fixed value of $\kappa$ and arbitrary feedback-phase $\varphi$. $\Delta\omega = \omega_s - \omega_0$ and $\Delta G = g'(N_s - N_0)$ in eq. (2.1) are the CCM optical frequency and threshold gain offsets from those of the solitary SL. MGM should have a smallest possible (negative) $\Delta G$, i.e., should satisfy the condition $\partial_{\Delta\omega} F = 0$, which implies $\Delta G = \frac{\tau_0 \Delta\omega}{\alpha_H l}$ and, thus, $\Delta\omega_{MGM} = -\frac{\alpha_H}{\tau_0} \ln\left(\frac{1+\kappa/|r_0|}{1+\kappa|r_0|}\right)$. From here we immediately get the MGM wavelength detuning from the corresponding solitary FP SL resonance,

$$\Delta\lambda_{MGM}(\kappa) = \frac{\lambda_0^2}{2\pi c} \frac{\alpha_H}{\tau_0} \ln\left(\frac{1+\kappa/|r_0|}{1+\kappa|r_0|}\right), \qquad \text{eqn. (2.2)}$$

and the related MGM frequency offset (or secondary StF) that is used in the rf domain,

$$\Delta\nu_{MGM}(\kappa) = \frac{\alpha_H}{2\pi\tau_0} \ln\left(\frac{1+\kappa/|r_0|}{1+\kappa|r_0|}\right). \qquad \text{eqn. (2.3)}$$

The corresponding relations in the L-K model are $F^{LK} = \left(\Delta\omega - \frac{l\alpha_H \Delta G}{\tau_0}\right)^2 + \left(\frac{l\Delta G}{\tau_0}\right)^2 - |\eta|^2$, $\Delta\omega_{MGM}^{LK} = -\alpha_H|\eta|$, and $\Delta\lambda_{MGM}^{LK}(|\eta|) = \frac{\lambda_0^2 \alpha_H |\eta|}{2\pi c}$, $\Delta\nu_{MGM}^{LK}(|\eta|) = \frac{\alpha_H |\eta|}{2\pi}$. Since $\eta$ and $\kappa$ are related by eq. (1.2) these L-K model-based relations provide perfect approximations of (2.2) and (2.3) when $\kappa$ is small.

Functions $\Delta\nu_{MGM}(\kappa)$ and $\Delta\lambda_{MGM}(\kappa)$ calculated for $\alpha_H = 3.5$ are represented by thin dashed curves in both panels of Fig. 3. One can see that the $\Delta\lambda_{MGM}(\kappa)$ curves fit the main spectral peak wavelengths of multiple FP SL resonances well (see Fig. 3(b)). Also, $\Delta\nu_{MGM}(\kappa)$ maps to secondary rf peak position at the upper border of the RF spectrum within the (0-70) GHz range (see Fig. 3(a)). However, the rf spectrum has very little power at these higher frequencies, and the primary StF peak in the rf spectrum occurs at about one third of $\Delta\nu_{MGM}$.

The rf spectrum (0 – 15 GHz range) is examined at higher resolution in fig. 4(e)-(h), for $\alpha_H$ = 3.0, 3.5, 4.0 and 5.0. The modulation of the rf power level on the scale of the CCM frequency spacing, shows as stripes spaced by ~220 MHz. The optical spectrum associated with the strongest solitary laser resonances is shown in fig. 4(a)-(d), with one or two SL resonances in each frame, for the same four values of $\alpha_H$. Simulations with $\alpha_H$ values of 2.0 or less give a dynamic mapping that is qualitatively different to that shown in fig. 4, with no clear transition in and out of the coherence collapse regime. Such results do not contribute to the discussion of generating outputs with a broad chaos bandwidth and are not included here. They do indicate that there is further LEF dependent diversity in the dynamics of the SLDOF system and that an $\alpha_H$ value of ~3 or greater is needed to have a well-defined coherence collapse region in the considered SLDOF.

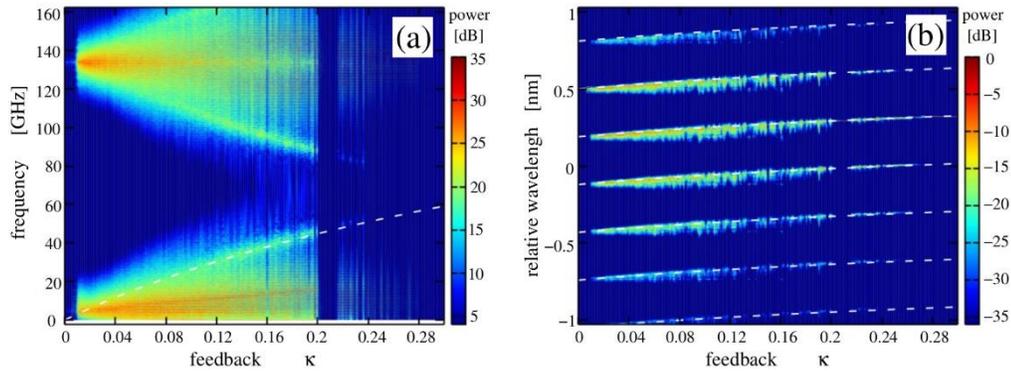

Fig. 3 (a) Rf power, and, (b) Optical spectrum mappings, as a function of amplitude optical feedback factor, κ, color coded on a dB scale. Dashed curves: functions $\Delta v_{MGM}(\kappa)$ and $\lambda_{MGM}(\kappa)$ calculated for $\alpha_H = 3.5$. Note the laser operates single frequency for κ < 0.011. 7 longitudinal modes are shown oscillating within the relative wavelength range [-1,1] nm after coherence collapse has occurred.

The results in fig. 4, including the overlaid dashed lines indicating the evolution of $\Delta\lambda_{MGM}(\kappa)$ in the upper panels and $\Delta v_{MGM}(\kappa)$ in the lower panels, are consistent with those introduced and discussed for fig. 3. Though, for $\Delta\lambda_{MGM}(\kappa)$, the dashed lines start on or above the wavelength with highest power for low κ values and switch to below it at higher κ values. The discrepancy is small and not unexpected given equations (2.2) and (2.3) used a simplified TW model. The primary StF rf peak frequency also grows with $\alpha_H$ and κ All the analyses reinforce the importance of having a high value of $\alpha_H$ for achieving a maximized chaos bandwidth. Also, as $\alpha_H$ increases the number of harmonics of the ROF that are within the bandwidth of the rf spectrum, and which are therefore undamped, increases. At $\alpha_H = 3$ only up to the second harmonic of the ROF is undamped within the main coherence collapse region. This increases to undamping of up to the fourth harmonic of the ROF with $\alpha_H = 5$, when the frequency range of fig 4(h) is extended to 25 GHz. Also, ROF sidebands on one of the strong laser resonances can be seen, close to the coherence collapse boundary, in fig 4(d) . The rf spectrum also has increased rf power as $\alpha_H$ increases. These simulation results show that as the feedback fraction and/or the LEF increases it is the StF increasing that leads to the increased chaos bandwidth. Within this bandwidth there may be broad peaks centered at the primary StF, and, the ROF. Possibly also the harmonics of the ROF that occur within the rf spectrum. More details on undamping of the ROF and its harmonics within the StF range will be given in the next subsection.

For κ values up to about 0.12 in the coherence collapse time-averaged optical spectrum there is reasonable, but varying, power across the full range from a solitary laser resonance to the associated MGM wavelength (fig. 4 (a)-(d)). Above this value of κ the wavelength range closest to the solitary laser resonance has less power. From an instantaneous wavelength/frequency perspective, this suggests the wavelength rarely revisits the solitary laser resonance, and remains closer to MGM wavelength for most of the time at these larger κ values. A consequence is that the full potential for the chaos bandwidth, as κ increases beyond 0.12, is not accessed.

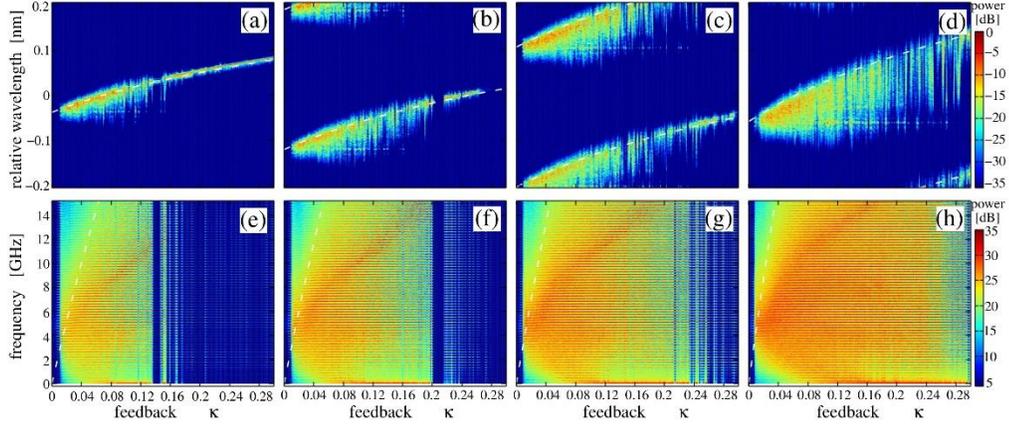

Fig. 4. Spectral mappings as functions of amplitude feedback factor κ for different values $α_H$, which are 3 in (a) & (e), 3.5 in (b) & (f), 4 in (c) & (g), and 5 in (d) & (h). -Top: portion of the optical spectrum containing at least one resonance of the solitary laser. Bottom: part of the rf Dashed curves: $λ_{MGM}(κ)$ (top) and $Δν_{MGM}(κ)$ (bottom) calculated for corresponding $α_H$.

The results in fig.4 show the expected reduction in the critical level of feedback required to trigger coherence collapse as $α_H$ increases. This feedback level is ~0.0097 at $α_H$=3, ~0.0088 at $α_H$=3.5, ~0.0079 at $α_H$=4 and ~0.0070 at $α_H$=5.

For the simulations represented in fig. 5, $α_H$ is 3.5, and it is the effect of changing the nonlinear gain compression parameter, ε, that is shown. The optical spectrum around one or two of the strongest SL resonances are shown in the top-row panel (fig. 5 (a)-(c)) and the rf spectra, for a frequency range of 0-15 GHz, are shown on the lower-row (fig. 5 (d)-(f)). The gain compression factor, ε, is varied over two orders of magnitude. The lowest value of $3×10^{-24}$ m$^3$ is shown in fig. 5 (a) and (d), the standard value of $3×10^{-23}$ m$^3$, used in the simulations presented so far, is in (b) and (e), and a high value of $3×10^{-22}$ m$^3$ is in (c) and (f). The increased damping of the relaxation oscillations within the rf spectrum of the coherence collapsed SLDOF, as ε increases, is dramatic. With a low ε value of $3×10^{-24}$ m$^3$ the regime III-IV transition to coherence collapse is heralded by strong undamping of both the relaxation oscillation and its second harmonic (see fig. 5(d)). The critical feedback level for the III-IV transition has been reduced to 0.0030 (from 0.0088 observed in fig. 5(e)). The maximum chaos bandwidth is increased by a factor of ~2. The power in the chaos spectrum is increased. The corresponding optical spectrum associated with one of the main-solitary-SL resonances (fig 5 (a)) indicates an increased spectral coverage between the SL resonance and the MGM wavelength, and also an enhanced blue-shifted section of the spectrum below the SL resonance wavelength. Increasing ε to $3×10^{-22}$ m$^3$ results in a dramatic suppression of the relaxation oscillation and a reduction in the chaos bandwidth by a factor of order 3 (cf ε = $3×10^{-23}$ m$^3$). The power in the rf spectrum is also significantly reduced, as is the ROF which is ~1.5 GHz. There is still evidence of enhancement in rf power through a rf range spanning several harmonics of the, now much lower ROF. For the low and high ε values in fig. 5 the transition to broad bandwidth coherence collapse regime occurs gradually with an increase of κ. It is with ε~$3×10^{-23}$ m$^3$ that exceeding feedback of κ~0.0011 the previously damped oscillations are immediately undamped across a large rf range.

The behavior of the ROF in the numerical simulations, with increasing gain compression factor, is a reasonable fit to previously published theory [42, 43]. This predicts the ROF will change little, with increasing ε, until it starts to decrease quite rapidly. For device parameters in [42], which are similar to those used in the simulations here-in, a value of ε ~$4×10^{-23}$ m$^3$ is where the ROF starts to roll off quite rapidly. Values between $2\times10^{-23}$ m$^3$ and $4\times10^{-23}$ m$^3$ were optimal for achieving a reasonably flat modulation transfer function without significant reduction in the modulation bandwidth in that study [42]. The rf bandwidth reduction at larger values of ε is caused by nonlinear gain suppression of the ROF value. Thus, the ROF behavior with ε, reported here-in, is consistent with expectation from prior research.

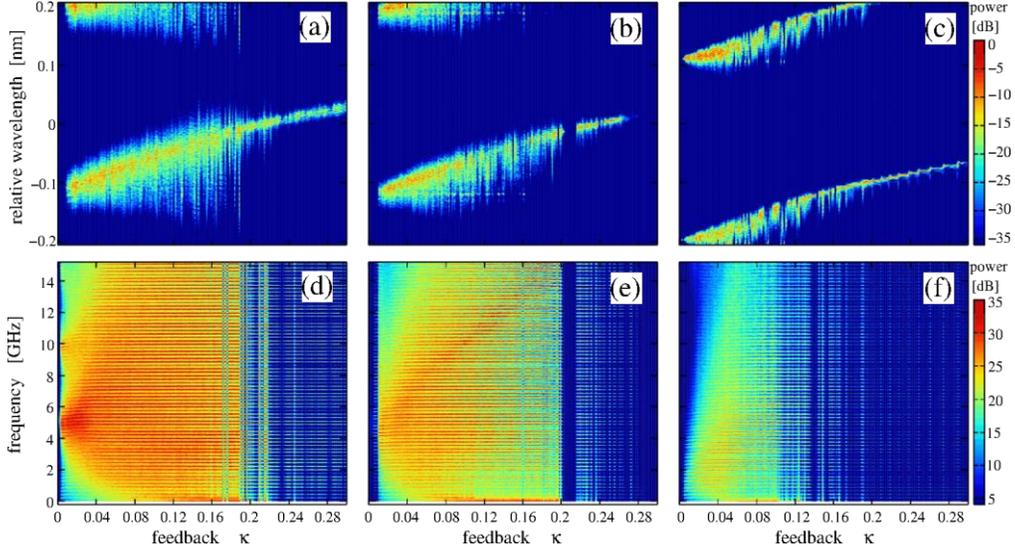

Fig. 5. Spectral mappings as functions of amplitude feedback factor κ for different values of the nonlinear gain compression factor ε, which are $3\times10^{-24}$ m$^3$ in (a)&(d), $3\times10^{-23}$ m$^3$ in (b)&(e), and $3\times10^{-22}$ m$^3$ in (c)&(f). $\alpha_H$ =3.5 in all cases. Top: portion of the optical spectrum containing at least one resonance of the SL. Bottom: part of the rf spectrum.

To summarize, the numerical simulations suggest a description of what is happening phenomenologically in the III-IV transition into coherence collapse. As the optical feedback factor κ is increased, the StF range reaches and/or includes the ROF and then the relaxation oscillations are undamped. A broadened rf spectrum, the chaos bandwidth, then spreads around the ROF. As κ increases further this bandwidth increases, spreading to both lower and higher frequencies than the ROF. It eventually reaches dc on the low side and can include up to several harmonics of the ROF on the higher frequency side. The commonly reported variation across the chaos bandwidth of 6-15 dB in the peaks in the average rf power level corresponds to enhancements occurring around the rf peak related to the StF, and, in some cases, frequencies close to the ROF and its harmonics. Increasing the LEF and/or decreasing the damping of the relaxation oscillation by, for example, decreasing the nonlinear gain compression factor, will increase the chaos bandwidth that is achieved at a given optical feedback factor in the coherence collapse region. When the lower bound of the rf spectrum reaches dc, the low frequency spectral component increases in power with further increase in κ. This means there will be an identifiable low frequency dynamic, such as low frequency fluctuations, occurring as part of the dynamical output. This is found to be consistent with experimental data as is discussed in section 2.3.

## 2.2 Numerical simulations - linear scale

As noted in the introduction, the use of a logarithmic scale for visualizing chaos bandwidth in the coherence collapse regime de-emphasizes the contrast to a standard 3dB bandwidth definition. Plotting the rf spectrum on a linear scale shows the variation in the peak heights of the spectral components and also shows clearly the CCM-spacing induced peaks which can have a 20 dB or greater variation in rf power. The discrete Fourier transform used for calculation of the rf spectra may introduce some inexactitudes but it is a consistent tool which allows the spectra to be compared. A selection of the rf spectra from fig. 3(a), with different values of κ, are plotted in fig. 6 for the frequency range 0-20 GHz on the left with an expansion of the range 3-7 GHz on the right. The latter shows the variation in the spectra at and around the ROF of ~5GHz.

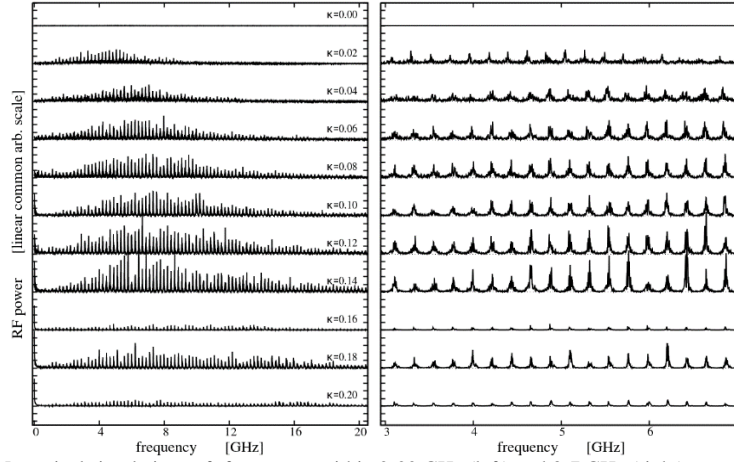

Fig. 6. Numerical simulations of rf spectrum within 0-20 GHz (left) and 3-7 GHz (right) ranges plotted on a common linear scale for feedback factor κ values indicated in the left panel.

Firstly, the increase in the primary StF, correlating with an increase in the chaos bandwidth with feedback factor, is seen in the left panel of fig. 6. However, a key feature of each chaos spectrum that is illustrated by fig. 6 relates to the power in the spectral features spaced at the CCM frequency separation ($\sim 1/\tau$). It is both the height and effective spectral coverage of these spectral features that should be considered when looking to define the optimal chaotic output that can be obtained from the SLDOF system. Larger effective spectral coverage is indicated by (i) larger full width at half maximum (FWHM) values, and/or, (ii) a lineshape with a broad pedestal, for the individual rf peaks. While the spectrum for κ=0.14 has the highest amplitude rf spectral features, it has a smaller effective spectral coverage because of smaller FWHM values for the individual peaks, than the CCM related features in the spectrum for κ=0.04. This indicates less overall spectral coverage within the chaos spectrum for the former. The chaos spectra up to ~κ=0.14 are suggested to have more spectral coverage on this measure than those for larger κ values.

As the feedback factor approaches the regime IV-V transition in this particular sequence of rf spectra, for κ=0.16 the rf power in the chaos bandwidth drops significantly, then recovers at κ=0.18, before dropping again just before the transition to regime V. We attribute this behavior to the stronger interaction of the primary StF with the third harmonic of the ROF for κ=0.18 than it has for κ=0.16. As the feedback factor increases further, the transition to regime V can be conceptualized as the frequency sweep not providing sufficient interaction with the ROF or any of its harmonics to support the continuance of coherence collapse. The mappings of corresponding fig. 3 at κ>0.22 do show an occurrence of transient unstable dynamics with low overall rf power, and single- or nearly single wavelength operation. The latter is characterized by a small-range-variation of carrier density, a mostly regular contribution of a single well defined MGM, and only a few weakly or moderately suppressed CCMs in its vicinity, and, possibly, in the vicinity of the MGM of a neighboring solitary FP resonance. The primary StF is approaching the fourth harmonic of the ROF, see enhancement of spectral peaks in fig. 3(a) at about κ=0.23 and 20 GHz frequencies, but the interaction is not sufficient to sustain coherence collapse as κ continues to increase.

The rf spectrum obtained by a filtering process, which smooths the CCM-induced peaks, illustrates the rf power and range covered, as referenced to the ROF and primary StF. Fig. 7 shows this on a linear rf power scale for frequency averaging intervals of (a) $1/\tau$, (b) $2/\tau$, and (c) $3/\tau$. At small κ values these show a broad peak at or near the ROF. A peak is seen at the second harmonic of the ROF when it appears to be coincident with the primary StF. The increasing bandwidth with increasing κ is an indicator of the StF. A reminder is given that this filtering process does imply a complete rf spectral coverage when the rf spectrum does have discrete CCM frequency spacing peaks.

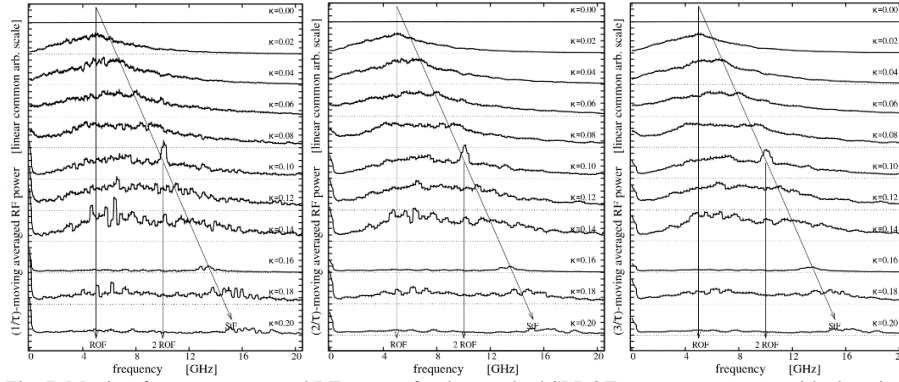

Fig. 7. Moving-frequency-averaged RF spectra for the standard SLDOF system parameters with changing feedback factor κ. (a) $(1/\tau)$-averaging, (b) $(2/\tau)$-averaging, (c) $(3/\tau)$-averaging.

## 2.3 Experimental results - linear scale

Experimental rf spectra (16 GHz detection bandwidth), calculated as the MATLAB fast Fourier transform of the full time series data [40] at an injection current of 65 mA, for several feedback factors are shown in fig.8. The feedback factors increase from top to bottom and their estimates are given in the figure caption along with the corresponding AOM voltage from the dataset. These are estimated values calculated as 0.0625 times the square of the measured normalized transmission of the AOM at a given voltage. The double pass through all elements in the external cavity is included by this. The coupling coefficient is not known accurately and has been estimated as 25% which is consistent with values measured for similar systems operated in our laboratory [44, 45]. Use of an aspheric lens for collimating the optical feedback is an improvement over the collimating lenses used in [44,45]. The ROF of the experimental SLDOF has been measured as ~1.5 GHz at 65 mA. This, and its integer multiples, are consistent with the frequencies at which enhancement is observed in the full rf spectra. In common with the numerical simulations the rf bandwidth has a maximum spread of about four times the ROF. But, because the ROF in the experiments is less than a third that in the simulations the maximum rf bandwidth is reduced by a similar fraction. The comparison of the experimental results in [22] with the simulations here-in indicates that the nonlinear gain compression factor for the APL device is likely larger than the standard value used here-in ($3\times10^{-23}\,\text{m}^3$) but not by as much as an order of magnitude. This explains the reduced ROF in the experiments cf the simulations and is discussed in more detail below. There is evidence of spectral peaks associated with CCM frequency spacing up to ~12 GHz in the experimental rf spectra for high feedback factors, but the power in these features is significantly reduced compared to that in the spectrum up to ~6 GHz. The graphs have been limited to a range of 12.5 GHz because there are no discernible spectral peaks above this frequency in the data. The 16 GHz data has captured the dynamics of the system. The important characteristics of height, width and fill of the CCM spaced features are in qualitative agreement with the simulations. It is both the height and width, and particularly the spectral coverage within the CCM spectral peak components that indicate how an optimal chaotic output might be defined.

Fig. 9 presents a set of 8 rf spectra from the 16 GHz bandwidth dataset (65 mA) (left and center) contrasted with those from the 4GHz bandwidth dataset (60 mA) (right). Note that for the 4 GHz data the roll off by 5 GHz is imposed by the measurement bandwidth. For the 16 GHz bandwidth data in this case, as compared to fig. 8 above, only every 4th data point of the fast Fourier transform has been recorded, in order to overcome practical limitations of the graphing package used. The feedback factors increase in value from top to bottom and their estimates are given in the figure caption. The ROF of the experimental SLDOF has been measured as ~1.5 GHz at 65 mA and ~1.2 GHz at 60 mA, both with an uncertainty of order 0.2 GHz. The key features described for fig. 8 are seen in the rf spectra of fig. 9 as well, with the exception that the spectral peaks for the 16 GHz bandwidth data would be more filled in, as seen in fig. 8, if all data points could have been plotted. The experimental rf spectra show the StF increasing with feedback factor and the power is enhanced in frequency bands near the ROF and its harmonics. The experimental results show the narrowing of the rf spectral peaks as the feedback factor increases more clearly than do the simulation results. This is particularly apparent when the detection bandwidth was insufficient to capture the dynamics fully as is the case for the 4GHz detection bandwidth.

The features that are now being proposed as important ones to consider in evaluating chaos in the SLDOF - all of the height, width and spectral fill of the spectral peaks of the rf spectrum – will now be linked with earlier mappings that have been done using the 4 GHz dataset as published in [22]. Informed by the results of the numerical simulations here-in, it is the root mean square (RMS) amplitude of the time series data that is identified as the best measure reported in the prior work ([22], fig. 2(a)) to decide an overall optimization of the chaotic output. RMS amplitude of the SLDOF time series output should be maximized. This occurred for feedback factors in the mid-range for the coherence collapse region for the 4GHz dataset, i.e. optical feedback factors values centered about ~0.125 referring to fig. 2(a) of [22]. This corresponds to the best tradeoff of having higher power in the time varying component of the output and greater spectral fill/coverage within the chaos. If the optimization for a specific application does not need the highest rf power in the chaotic output then lower optical feedback factors may give more complete rf spectral coverage within a sub-maximal chaos "bandwidth". These are issues which can be investigated in future experimental studies in any semiconductor laser or semiconductor amplifier based chaotic system.

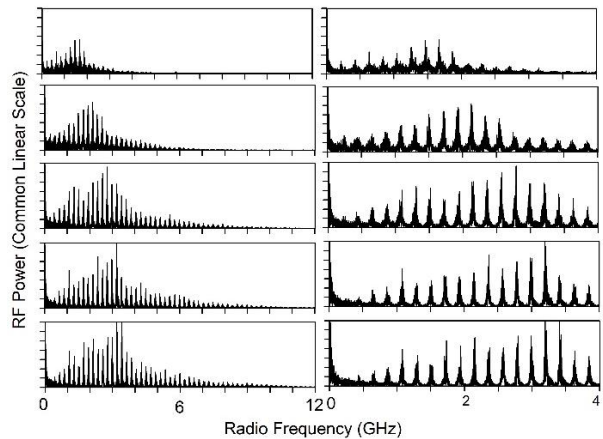

Fig. 8. Experimental data (50,000 points, sampling period 20 ps), 16 GHz detection bandwidth, rf spectrum 0-12.5 GHz (left), with expansion of 0-4 GHz (right), plotted on a common linear rf power scale, for estimated amplitude feedback factors (from top to bottom) of 0.0564, 0.0795, 0.107, 0.123, 0.126 (AOM voltages 0.650 V, 0.610 V, 0.570 V, 0.550V, 0.546 V; respectively). Regime III to IV transition is at AOM voltage 0.740 V (0.0275) and regime IV to V transition is at 0.542 V (0.1286). Injection current 65 mA.

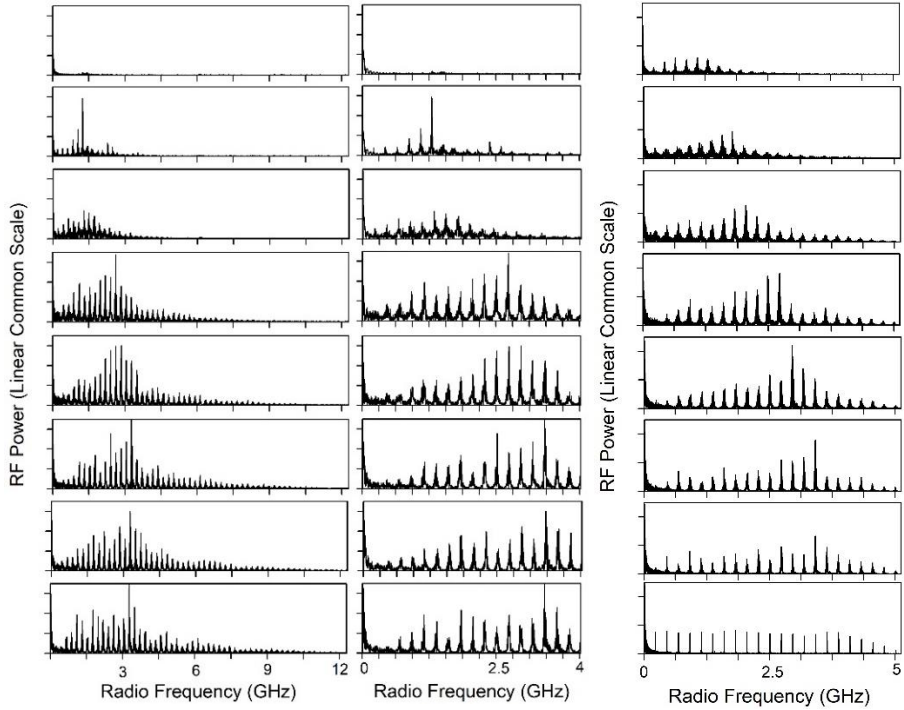

Fig. 9. Experimental data, 16 GHz detection bandwidth, rf spectrum 0-12.5 GHz (left), with expansion of 0-4 GHz (center) plotted on a common linear scale for relative amplitude feedback factor values (from top to bottom) of 0,0260, 0.0362, 0.0563,0.0795, 0.0939, 0.1228, 0.1257, 0.1272. (AOM voltages: 0.750V, 0.700 V, 0.650 V, 0.610 V, 0.590 V, 0.550 V, 0.546 V, 0.544 V) Injection current is 65 mA (~1.5 $I_{th}$). Note only every 4th data point of the calculated rf spectrum has been plotted as cf fig. 7 where every data point is plotted). 4 GHz detection bandwidth, rf spectrum 0-5 GHz (right) for relative feedback factor values (top to bottom) of 0.0353, 0.0534, 0.0777, 0.122, 0.138, 0.153, 0.167, 0.168 (AOM voltages: 0.680 V, 0.640 V, 0.600 V, 0.540 V, 0.520 V, 0.500 V, 0.482 V, 0.480 V). Injection current is 60 mA (~1.4 $I_{th}$).

In connecting with other systems that have been studied experimentally in the past we note that fig 5(c) is a good match for the rf spectra from a SLDOF system using an STC LT50-03U SL [46]. This system operated in the coherence collapse regime with no measurable evidence of the relaxation oscillation showing in the optical and rf spectra. The latter were measured with a rf spectral analyzer. Regime IV showed rf spectral features associated with the mixing of CCMs only. This high damping of the relaxation oscillations for the STC device motivated a new measurement method to measure the ROF [47]. The results of [46] can now be interpreted as coming from a system using a SL with a higher nonlinear gain compression factor than that of the APL-830-40 FP SL-based system [22, 38, 39], something of the order of $10^{-22}\,m^3$.

## 2.4 Further discussion

The results presented here-in indicate that the preferred SL parameters for achieving optimal chaotic output when operated in a SLDOF system are different to the standard parameters of commercial laser devices that have been driven by communications applications, for example. The drivers for commercial SL design have sought decreased LEF, increased ROF for a given injection current with sufficient damping of the RO to achieve a broad, flat modulation bandwidth. Broad bandwidth chaotic output from the coherence collapse regime of a SLDOF system will be enabled by using a SL with a high value for the LEF and with low damping of the RO. It may also prove to be advantageous to have a modest ROF if the system can be designed to undamp many higher harmonics of the RO by the optical feedback. For the experimental system analyzed above the main chaos bandwidth covers up to 4×ROF (~6 GHz) with reduced rf power CCM-derived spectral components at up to 8×ROF (~12 GHz). This latter observation is consistent with numerical simulations which show a second, lower power component to the rf spectrum between the primary StF and secondary StF. The general SL parameters for good SLDOF chaotic output being reported herein will be relevant to devices used in other nonlinear systems, as introduced above [10-18], where chaotic bandwidth without the imprint of the CCM-interaction-induced-frequencies can be achieved. It would be fair to say that almost all SLs used in SLDOF studies to date have been sub-optimal for the application due to the use of commercial SLs, or devices fabricated to conform to standard practice for commercial devices, as used in optical communications.

All experimental measurements of the optical frequency spectrum (not presented in this work) are time averaged relative to the fast timescales of the output power dynamics of the SLDOF. However, the instantaneous wavelength (and frequency) are changing dynamically. The range of wavelength between an SL resonance and its MGM wavelength, as optical feedback is increased, is identified as a measurement that should be investigated experimentally to further test the predictions of the numerical simulations presented here-in. Experimental equivalents of fig. 3(b) can be generated. Measurement of the change in wavelength of each SL resonance that is lasing, as a function of the optical feedback factor, has not been carefully tracked in most experimental studies to date. It is a recommendation for future studies. Measuring the instantaneous frequency (or wavelength) would be of value if a method to do so was available. In TW-model based simulations of chaotic SLDOF systems, the instantaneous frequencies are accessible through the mode analysis of the calculated fields [33, 48]. Such analysis, however, is time consuming for the considered system containing several thousand possible CCMs of importance and, thus, will be discussed elsewhere. The time averaged optical frequency spectrum that is measured can be interpreted as showing which wavelengths the device spends time at. As the optical feedback increases the wavelength appears to spend a decreasing amount of time in the vicinity of any given SL resonance from the numerical simulations. This appears to correlate with the predicted rf bandwidth of the chaotic regime being about one third of the full MGM wavelength offset range, when the latter is expressed in frequency units. The importance of this detail of the time averaged optical frequency spectrum, and how it scales with feedback factor, LEF and nonlinear gain compression factor, and, the connection with the rf spectrum, has been shown in this work. It emerges that SLDOFs should be designed that maximize the MGM wavelength offset, which can be done by increase of LEF, while aiming for the instantaneous output to spend a relatively uniform amount of time at all wavelengths in the range between this and the free running resonance wavelength of the SL. This will be advantageous to achieve a chaos bandwidth with relatively uniform power throughout.

When considering the optical frequency spectrum around a single resonance of the solitary FP SL, as shown in figs. 4(b) and 5(b) for the standard model parameter values, the loss of connection of the MGM wavelength to the solitary laser resonance wavelength is seen for feedback factors in the range 0.15-0.20 which are within the main regime IV coherence collapse region. This also heralds when the individual CCM-based features in the rf spectrum start to show narrower peaks. We suggest that the chaotic output is no longer optimal in this region for this reason, even though the chaos bandwidth continues to increase. This raises the question of how the rf spectral coverage across the full MGM wavelength offset range might be sustained to higher optical feedback factors. Optoelectronic feedback [2-4], frequency shifted optical feedback (FSF) [49, 50], and electro-optically frequency-modulated optical feedback (FM

SLDOF) [51, 52] are all technical approaches which support a more uniform and complete spectral coverage of the instantaneous wavelength within the range between the solitary resonance and the MGM wavelength over time. The system with optoelectronic feedback has been accepted as a good source for broadband optical chaos. The FSF SLDOF and FM SLDOF are worth reconsideration in light of the new insight into the importance of achieving more uniform and complete spectral coverage within the full MGM wavelength offset range.

## 3. Conclusion

Numerical simulations of a SLDOF system with a focus on the rf spectrum and optical spectrum of the output in the main regime IV coherence collapse region have been completed. These have uncovered the importance of the optical spectra within the MGM wavelength offset range. Eqn. (2.2) describes the dependence of this range on feedback strength $\kappa$ and LEF. This dependence has been supported by the numerical simulation results reported herein. The primary StF rf peak in the broad-band rf spectrum is located at about one third of the MGM frequency offset $\Delta\nu_{MGM}(\kappa)$ defined in eq. (2.3). The undamping of the RO and its harmonics, within the primary StF range drives oscillation in and around the CCMs associated with each SL resonance. The nonlinear mixing of the CCM-beating-induced oscillations then leads to rf spectral peaks that are broad and noisy when the chaos is optimum. Using a SL with low damping of the relaxation oscillation, which can be facilitated, e.g., by having a low nonlinear gain compression factor for the SL, will support increased chaos bandwidth. Also, a SL with a high LEF should be used. The optimal rf spectrum of the chaotic output will depend on the end use. If maximizing the peak to peak variation in the chaotic output is required then optimal operation will be for mid-range optical feedback levels within the coherence collapse region. The RMS amplitude of the SLDOF output is found to be an excellent measure for identifying this optimized region. It should be maximized. If an application can be serviced by a lower overall rf power, smaller bandwidth chaotic output, then operating closer to the regime III-IV transition boundary will be preferred. The maximal bandwidth chaotic output obtained at the regime IV-V transition boundary is to be avoided as the CCM-mixing-based rf components narrow as this boundary is approached even though the bandwidth continues to increase. Experimental evidence supports these conclusions. This increases confidence that numerical artefacts that could affect the rf spectra obtained from the numerical simulations are not qualitatively impacting the predictions.

It has been established in prior research that the SLDOF is not the preferred SL-based nonlinear system for generating broadband chaos, in several applications, due to the strong imprint of the CCM frequency spacing in the rf spectrum. But based on the results presented here-in SLDOF systems could be much improved in terms of their chaotic output if they use appropriately designed SLs. Also, optoelectronic feedback, frequency shifted feedback, and intra-EC electro-optic modulation (FM) are all technical approaches that can be revisited for achieving improvements in chaotic bandwidth, and rf spectral coverage within that bandwidth, if a more optimal SL is used. A more chaotic output relative to the basic SLDOF has been previously reported in these systems.

**Funding**

**Acknowledgments.** Dr Joshua Toomey is acknowledged and thanked for completing experiments generating much of the data in the publicly available datasets [40], two of which have been used in this work. The team at Securities Industry Research Centre of Asia-Pacific (SIRCA), Sydney, Australia, are acknowledged for creating the database as part of a Science and Industry Endowment Fund (SIEF) funded project - http://laser.portal.knowledgediscovery.org/dataset. DMK created the dataset https://doi.org/10.6084/m9.figshare.c.3834916.v1. This work has not had specific project funding but it has benefited from prior funding. MR was funded by a Macquarie University Faculty of Science and Engineering Visiting Researcher Fellowship, which facilitated the initiation of this collaboration. Generation of the publicly available experimental datasets that have been used in this research was supported by an Australian Research Council (ARC) Linkage Project LP100100312 and a SIEF grant – RP-04-174

**Disclosures.** The authors declare no conflicts of interest.

**Data availability statement.** Experimental data underlying results presented in this paper are available as given in Ref. [40] and in the acknowledgments. Data generated by the numerical simulations is not publicly available. It is generated as needed and is, in general, not permanently stored. Queries about the simulations and simulation data should be sent to MR, radziuna@wias-berlin.de.